\documentclass[aps,prd,twocolumn,groupedaddress,%
amsmath,amssymb,eqsecnum,showpacs,showkeys]{revtex4}
\usepackage{graphicx}

\providecommand\half{\tfrac{1}{2}}
\providecommand\thrhlf{\tfrac{3}{2}}

\providecommand{\abst}[1]{\lvert#1\rvert}
\providecommand{\absd}[1]{\left\lvert#1\right\rvert}
\providecommand{\bra}[1]{\langle#1\rvert}
\providecommand{\ket}[1]{\lvert#1\rangle}
\providecommand{\braket}[2]{\langle#1|#2\rangle}

\DeclareMathOperator{\sech}{sech}

\DeclareMathOperator{\Arccosh}{Arccosh}
\DeclareMathOperator{\Arcsinh}{Arcsinh}

\begin{document}

\title{Localized Particle States and Dynamic Gravitational Effects}

\author{Ian H.~Redmount}

\email{redmouih@slu.edu}

\affiliation{Department of Physics\\
Parks College of Engineering, Aviation, and Technology\\
Saint Louis University\\
St.~Louis, Missouri~~63103--1110~~USA}

\date{\today}

\begin{abstract}
Scalar particles---i.e., scalar-field excitations---in de~Sitter space exhibit
behavior unlike either classical particles in expanding space or quantum
particles in flat spacetime.  Their energies oscillate forever, and their
interactions are spread out in energy.  Here it is shown that these features
characterize not only normal-mode excitations spread out over all space, but
localized particles or wave packets as well.  Both one-particle and
coherent states of a massive, minimally coupled scalar field in de~Sitter
space, associated with classical wave packets, are constructed explicitly.
Their energy expectation values and corresponding Unruh-DeWitt detector
response functions are calculated.  Numerical evaluation of these quantities
for a simple set of classical wave packets clearly displays these novel
features.  Hence, given the observed accelerating expansion of the Universe,
it is possible that observation of an ultralow-mass scalar particle could
yield direct confirmation of distinct predictions of quantum field theory
in curved spacetime.
\end{abstract}


\maketitle

\section{INTRODUCTION\label{sec1}}

Although quantum field theory in curved spacetime is widely regarded as
a mature, even a finished field, experimental or observational
confirmations of its predictions are lacking.  It is therefore of interest
to seek phenomena within the theory---entailing both field quantization
and nontrivial spacetime geometry---which might come within reach of
actual measurement.  Effects of quantized fields on the evolution of
spacetime, e.g., via black-hole evaporation~\cite{hawk74} or cosmological
inflation~\cite{guth81}, have been extensively studied for decades, but
direct observation of either phenomenon has proved elusive.  It is possible
to look in the other direction:  for effects of spacetime geometry---in
particular, of spacetime \emph{dynamics}---on quantized-field excitations,
i.e., on the measurable properties of elementary particles.  The possibility
that precise measurements of particle properties might probe the dynamics
of the cosmos is an intriguing one.

In 1989 I described the behavior of scalar particles, i.e., excitations of
a real, linear, scalar field, in de Sitter space~\cite{redm89b}.  Modulation
of the field normal modes by the vigorous dynamics of that spacetime causes
those excitations to behave unlike either classical particles in an
expanding universe or quantized-field excitations in a static spacetime:
Their energy expectation values, above that of the vacuum, oscillate at
late and early times with fixed amplitude and frequency, rather than
redshifting to a constant value (the field mass).  And their interactions
with a ``detector'' coupled to the field, again above the vacuum value,
describe not a finite transition \emph{rate} at fixed energy but a finite
transition \emph{probability} over all time, with finite width in energy.
That is, the detector fails to fix the energy of an excitation at a
specific value.  This effect is also found for scalar-field excitations
in a suitably dynamic ``laboratory'' setting in flat spacetime~\cite{daue93}.

At the time, however, these phenomena might have been regarded as
arcane curiosities, for several reasons:  The expansion of an ordinary
matter- or radiation-dominated universe is far less vigorous than the
hyperbolic or exponential expansion of de~Sitter space, suppressing the
effects.  And these effects are apparent only for excitations of a field
with Compton wavelength comparable to the curvature radius of the spacetime,
being strongly suppressed for larger masses.  In the early universe this
might encompass fields with GUT-scale masses, but that epoch is inaccessible
to direct observation; in the present era this would require a mass
far smaller than that of any known particle.  (Conformally invariant
\emph{massless} fields, such as that of the photon, do not exhibit such
effects at all.)  Also, the normal modes of the field correspond to
field configurations extending over all space; observed particles are
associated with localized wave packets.

Recent discoveries, however, have vitiated some of these objections.  Detailed
observations of Type~Ia supernova distributions~\cite{perl98} and of the cosmic
microwave background suggest that the present Universe is dominated not by
matter or radiation but by cosmological-constant, vacuum, or ``dark energy''
contributions, making de~Sitter expansion a better approximation to
cosmological dynamics than previously apparent.  Moreover, Parker and
Raval~\cite{park99a,park99b} have shown that the vacuum-energy effects
of a linear scalar field with mass of order $10^{-33}$~eV can account
quite tidily for the observed expansion of the Universe.  Such a mass
corresponds to a Compton wavelength of order $10^{26}$~m, the same order
as the curvature radius of a de~Sitter-like spacetime with Hubble parameter,
say, 75~km/s/Mpc.  If the particles (excitations) of Parker's and Raval's
proposed field can be detected, they might exhibit the dynamical effects
I have described.  Burko, Harte, and Poisson~\cite{burk02} have also shown
that a \emph{classical} scalar particle in an expanding universe can display
unusual behavior---a nonconstant mass---with potentially observable
consequences.

In this work I show that the novel dynamical features of normal-mode
excitations of a scalar field in de~Sitter space can persist in field
states describing localized wave packets, e.g., first-quantized or
classical particles.  I construct several classes of (second-quantized)
field states associated with classical solutions of the field equation.
I then obtain exact expressions for excitation energies and monopole-detector
response functions in these states.  Numerical evaluation of these results
for a simple choice of wave packet illustrates the effects sought, and
suggests further avenues of exploration.

The spacetime geometry and field theory used here are described in
Sec.~\ref{sec2} below.  Quantum states suitable for describing localized
field configurations are given explicitly in Sec.~\ref{sec3}, and their
energies and detector responses are calculated.  Numerical results for
sample wave packets are shown in Sec.~\ref{sec4}.  The results and their
implications are summarized in Sec.~\ref{sec5}.  Throughout I use units
giving $\hbar=c=1$; sign conventions and general notation follow those
of Ref.~\onlinecite{mtw73}.   

\section{SPACETIME AND FIELD\label{sec2}}

\subsection{de~Sitter-space geometry\label{sec2a}}

de~Sitter space provides a suitably dynamic background geometry.
This spacetime can be given coordinates with closed (spherical),
flat, or open (hyperbolic) spatial sections.  Although recent evidence
suggests the latter two may be more appropriate to the actual
Universe~\cite{garn98,perl98}, I use the first here for simplicity.
For $N+1$-dimensional de~Sitter space (a straightforward generalization
of $3+1$~dimensions), convenient coordinates are comoving-observer
proper time $t\in(-\infty,+\infty)$, $N-1$~polar angles
$\theta_1,\ldots,\theta_{N-1}\in[0,\pi]$, and one azimuthal angle
$\phi\in[0,2\pi)$.  The spacetime metric takes the form
\begin{equation}
\label{eq01}
ds^2=-dt^2+a^2\cosh^2(t/a)\,d\Omega_N^2
\end{equation}
in these coordinates, where $a$ is a positive constant and $d\Omega_N^2$
the metric of the unit $N$-sphere.

\subsection{Scalar field theory\label{sec2b}}

A real, linear, massive, minimally coupled, quantized scalar field serves
as a simple example.  The field~$\varphi$, with mass~$\mu$, has Lagrangian
density
\begin{equation}
\label{eq02}
\mathcal{L}=-\half(\nabla_\alpha\varphi\,\nabla^\alpha\varphi+\mu^2\varphi^2)
\end{equation}
and corresponding field equation
\begin{equation}
\label{eq03}
(\square-\mu^2)\,\varphi=0\ ,
\end{equation}
where $\square$ is the covariant d'Alembertian in $N+1$-dimensional
de~Sitter space.

Quantized fields in de~Sitter space have been studied extensively since the
1950's~\cite{gutz56}. \ A thorough description of the approach used here,
with references to some of the enormous literature, is given in
Ref.~\onlinecite{redm89b}.  In this section I briefly summarize the
features of this model appropriate to the present problem.

\subsubsection{Canonical quantization\label{sec2b1}}

Canonical quantization of this field is effected via an expansion of~$\varphi$
in normal modes, with operator coefficients, thus:
\begin{equation}
\label{eq04}
\varphi(t,\Omega_N)=a^{-N/2}\sum_L[b_L\chi_L(t)+b_L^\dagger\chi_L^*(t)]
\,\mathcal{Y}_L(\Omega_N)\ .
\end{equation}
Here $\Omega_N$ denotes the angular coordinates $\theta_1,\ldots,
\theta_{N-1},\phi$ collectively, and $L$ denotes the set of angular-momentum
quantum numbers identifying the normal modes.  The $N$-dimensional spherical
harmonics~$\mathcal{Y}_L$ can be given explicitly as
\begin{subequations}
\label{eq05}
\begin{equation}
\label{eq05a}
\begin{aligned}[b]
\mathcal{Y}_L(\Omega_N)&=\prod_{j=1}^{N-1}\Biggl[\left(\lambda_j
\frac{(\lambda_j-\half+\lambda_{j+1})!}{(\lambda_j-\half-\lambda_{j+1})!}
\right)^{1/2}\\
&\qquad\qquad\times(\sin\theta_j)^{(j+1-N)/2}\,
\mathrm{P}_{\lambda_j-(1/2)}^{-\lambda_{j+1}}(\cos\theta_j)\Biggr]\\
&\qquad\qquad\qquad\times[\pi(1+\delta_{l_N0})]^{-1/2}\begin{Bmatrix}
\cos(l_N\phi)\\ \sin(l_N\phi)\end{Bmatrix}\ ,
\end{aligned}
\end{equation}
with
\begin{equation}
\label{eq05b}
\lambda_j\equiv l_j+\frac{N-j}{2}
\end{equation}
and
\begin{equation}
\label{eq05c}
l_1\ge l_2\ge\cdots\ge l_{N-1}\ge l_N\ge0
\end{equation}
\end{subequations}
the integer quantum numbers making up the set~$L$.  Here the P's are
associated Legendre functions of the first kind, and the harmonics are
chosen real for convenience.  The time-dependence functions~$\chi_L$ can
be written
\begin{subequations}
\label{eq06}
\begin{equation}
\label{eq06a}
\begin{aligned}[b]
\chi_L(t)&=\left(\frac{a\pi/2}{\sinh(\pi qa)}\right)^{1/2}\cosh^{-N/2}(t/a)\\
\times\Bigl\{\kappa_L^{(+)}&\mathrm{P}_{\lambda_1-1/2}^{-iqa}[\tanh(t/a)]
+\kappa_L^{(-)}\mathrm{P}_{\lambda_1-1/2}^{+iqa}[\tanh(t/a)]\Bigr\}\ ,
\end{aligned}
\end{equation}
with parameter
\begin{equation}
\label{eq06b}
q\equiv\left(\mu^2-\frac{N^2}{4a^2}\right)^{1/2}
\end{equation}
\end{subequations}
and constants $\kappa_L^{(\pm)}$ satisfying
$\abst{\kappa_L^{(+)}}^2-\abst{\kappa_L^{(-)}}^2=1$.  The form and behavior of
the functions differ for real and imaginary values of~$q$.  Here I shall take
$q$ real, i.e., mass satisfying $\mu>N/(2a)$.

The operator coefficients~$b_L$ and~$b_L^\dagger$ in expansion~\eqref{eq04}
play the usual role of field-excitation annihilation and creation operators,
respectively.  They satisfy commutation relations
$[b_L,b_{L^\prime}^\dagger]=\delta_{LL^\prime}$, \emph{et cetera.}

\subsubsection{Functional Schr\"{o}dinger description\label{sec2b2}}

Quantum states of the field~$\varphi$ can also be described by wave
functionals $\Psi[\varphi(\Omega_N),t]$, which depend on the configuration
of the field on constant-time hypersurfaces of the de~Sitter geometry, and
the time.  The wave functionals are solutions of a functional Schr\"{o}dinger
equation~\cite{free85,ratr85}
\begin{subequations}
\label{eq07}
\begin{equation}
\label{eq07a}
i\frac{\partial}{\partial t}\Psi=H\,\Psi\ ,
\end{equation}
with Hamiltonian operator
\begin{equation}
\label{eq07b}
H=\int d\Omega_N\,a^N\cosh^N(t/a)\,|g_{tt}|^{-1/2}T_{tt}
\end{equation}
constructed from the spacetime metric and the canonical stress-energy tensor
\begin{equation}
\label{eq07c}
T_{\alpha\beta}=\nabla_\alpha\varphi\,\nabla_\beta\varphi+
g_{\alpha\beta}\mathcal{L}
\end{equation}
\end{subequations}
of the field, where $\mathcal{L}$ is the Lagrangian density~\eqref{eq02}.
(For the minimally coupled field~$\varphi$, this tensor coincides with the
gravitational stress-energy tensor obtained by variation of the field action
with respect to the metric.)  The Hamiltonian is rendered a functional
differential operator via the representation of the field momentum
\begin{equation}
\label{eq08}
\Pi=\nabla_t\varphi\to ig_{tt}|g|^{-1/2}\frac{\delta}{\delta\varphi}\ ,
\end{equation}
with $g$ the metric determinant, paralleling the representation of momentum
in ordinary quantum mechanics.

The Hamiltonian takes a particularly simple form in terms of the amplitudes
with which the field configuration is expanded in spherical harmonics.  With
expansion
\begin{equation}
\label{eq09}
\varphi(\Omega_N)=a^{(1-N)/2}\sum_Ly_L\,\mathcal{ Y}_L(\Omega_N)
\end{equation}
---an expansion of the field on a single constant-time hypersurface, not on
the entire spacetime as in Eq.~\eqref{eq04}---the Hamiltonian becomes
\begin{subequations}
\label{eq10}
\begin{equation}
\label{eq10a}
H=\sum_L\left(-\frac{\sech^N(t/a)}{2a}\,\frac{\delta^2}{\delta y_L^2}+
\frac{a\cosh^N(t/a)}{2}\,\omega_L^2\,y_L^2\right)\ ,
\end{equation}
a sum of independent ``harmonic oscillator'' Hamiltonians with time-dependent
``angular frequencies''
\begin{equation}
\label{eq10b}
\omega_L=\left(\mu^2+\frac{\ell_1(\ell_1+N-1)}{a^2\cosh^2(t/a)}\right)^{1/2}\ .
\end{equation}
\end{subequations}
The Hamiltonian can also be expressed in terms of the amplitude operators and
time-dependence functions of expansion~\eqref{eq04}.  It then becomes
\begin{equation}
\label{eq11}
\begin{aligned}[b]
H&=\frac{\cosh^N(t/a)}{2}\sum_L\biggl[(b_L^\dagger b_L
+b_Lb_L^\dagger)(\dot\chi_L\dot\chi_L^*+\omega_L^2\chi_L\chi_L^*)\\
&\qquad\qquad\quad+b_L^2(\dot\chi_L^2+\omega_L^2\chi_L^2)
+b_L^{\dagger2}(\dot\chi_L^{*2}+\omega_L^2\chi_L^{*2})\biggr]\ ,
\end{aligned}
\end{equation}
where overdots denote time derivatives.  The annihilation operators are
represented in the functional Schr\"{o}dinger description as
\begin{equation}
\label{eq12}
b_L=-ia^{1/2}\cosh^N(t/a)\dot\chi_L^*y_L+a^{-1/2}\chi_L^*\frac{\delta}{
\delta y_L}\ ,
\end{equation}
with the creation operators $b_L^\dagger$ the Hermitian conjugate of this.

\subsubsection{Excitation-number eigenstates\label{sec2b3}}

A Fock space of quantum states for the field~$\varphi$ is spanned by
eigenstates of the excitation-number operators~$b_L^\dagger b_L$.  These
have wave functionals
\begin{subequations}
\label{eq13}
\begin{equation}
\label{eq13a}
\begin{aligned}[b]
\Psi_{\{n_L\}}[\{y_L\},t]&=\prod_L\left(\frac{a}{2\pi}\right)^{1/4}
(2^{n_L}n_L!)^{-1/2}\\
\times&[\chi_L^*(t)]^{-1/2}\left(\frac{\chi_L(t)}{\chi_L^*(t)}\right)^{n_L/2}
H_{n_L}[\Delta_L^{1/2}(t)\,y_L]\\
&\qquad\times\exp\{-\half[\Delta_L(t)-i\bar\Delta_L(t)]\,y_L^2\}\,
\end{aligned}
\end{equation}
with
\begin{align}
\label{eq13b}
\Delta_L(t)&=\frac{a}{2\abst{\chi_L(t)}^2}\ ,\\
\label{eq13c}
\bar\Delta_L(t)&=\frac{a\cosh^N(t/a)}{2\abst{\chi_L(t)}^2}\,\frac{d}{dt}
\abst{\chi_L(t)}^2\ ,
\end{align}
\end{subequations}
and each $H_{n_L}$ an Hermite polynomial of order~$n_L$.  The nonnegative
integral excitation numbers~$n_L$ for all the normal modes of the field
are exact constants of the states' evolution; the set~$\{n_L\}$ identifies
each eigenstate.  These are not, however, eigenstates of the
Hamiltonian~\eqref{eq10a} or~\eqref{eq11}.  These states form
an orthonormal basis:
\begin{equation}
\label{eq14}
\begin{aligned}[b]
\braket{\{n_L\}}{\{n_L^\prime\}}&=\int\Psi_{\{n_L\}}^*[\{y_L\},t]
\,\Psi_{\{n_L^\prime\}}[\{y_L\},t]\,\prod_Ldy_L\hfil\\
&=\prod_L\delta_{n_Ln_L^\prime}
\end{aligned}
\end{equation}
in bra-and-ket notation.

The expectation value of the field~$\varphi$ in any of these eigenstates
is zero.  Each wave functional~\eqref{eq13a} is of definite parity in each
field amplitude~$y_L$, implying
\begin{equation}
\label{eq15}
\begin{aligned}[b]
\bra{\{n_L\}}\varphi\ket{\{n_L\}}(t,\Omega_N)&=\int\Psi_{\{n_L\}}^*[\{y_L\},t]\\
\times\biggl(a^{(1-N)/2}\sum_Ly_L\,&\mathcal{ Y}_L(\Omega_N)\biggr)
\Psi_{\{n_L\}}[\{y_L\},t]\,\prod_Ldy_L\\
&=0\ .
\end{aligned}
\end{equation}
Of course the same result follows from expansion~\eqref{eq04}, the
annihilation and creation properties of the amplitude operators, and
the orthonormality relation~\eqref{eq14}.

The expectation value of the field energy---as defined by
Hamiltonian~\eqref{eq10a} or~\eqref{eq11}---in an excitation-number
eigenstate consists of a ``vacuum energy,'' plus a sum of excitation energies
for each mode times the number of excitations in that mode.  This can be
written
\begin{subequations}
\label{eq16}
\begin{equation}
\label{eq16a}
\bra{\{n_L\}}H\ket{\{n_L\}}(t)=\bra{\{0\}}H\ket{\{0\}}+\sum_Ln_L\,E_L(t)\ ,
\end{equation}
with
\begin{equation}
\label{eq16b}
E_L(t)=\frac{a^2\omega_L^2\cosh^{2N}(t/a)+\Delta_L^2+{\bar{\Delta}}_L^2}{
2a\Delta_L\cosh^N(t/a)}\ .
\end{equation}
\end{subequations}
The vacuum contribution~$\bra{\{0\}}H\ket{\{0\}}$, sometimes
described as a ``particle content,'' has been extensively
analyzed~\cite{dowk76,bunc78,birr82}.  It is the energy~$E_L$, however,
which is associated with a ``particle'' in the sense of an excitation
of the field.  For a minimally coupled field (specifically, for any
except a massless, conformally coupled field), $E_L$ behaves quite
unlike a classical energy.  For mass $\mu>N/(2a)$ as assumed here,
it oscillates at late times~$t>>a$:
\begin{subequations}
\label{eq17}
\begin{equation}
\label{eq17a}
E_L(t)\sim\frac{\mu^2}{q}\,\cosh\alpha_L+\frac{N\mu}{2qa}\,\sinh\alpha_L\,
\cos(2qt-\beta_L)\ ,
\end{equation}
where the parameters~$\alpha_L$ and~$\beta_L$ are determined by the
constants~$\kappa_L^{(\pm)}$ which fix the normal-mode time-dependence
(``positive-frequency'') functions~$\chi_L$, as in Eq.~\eqref{eq06a}, viz.,
\begin{align}
\label{eq17b}
\alpha_L&=2\Arccosh\abst{\kappa_L^{(+)}}=2\Arcsinh\abst{\kappa_L^{(-)}}\ ,\\
\label{eq17c}
\beta_L&=\arg\left[\kappa_L^{(+)}\kappa_L^{(-)*}\left(\frac{N}{2}+iqa\right)
\right]-2\arg\Gamma(1+iqa)\ .
\end{align}
\end{subequations}
The oscillations of the energy~$E_L$ do not damp out; their amplitude is
asymptotically constant.  For the minimally coupled field considered here,
the oscillating $E_L$ remain always positive.

Field states can also be characterized by the response of an Unruh-DeWitt
``monopole detector''~\cite{unru76,dewi79} coupled linearly to the field.  The
probability of the detector making a transition of energy~$E$---with the field
in state~$\ket{\{n_L\}}$---is proportional to a response function independent
of the detector's structure, viz.,
\begin{equation}
\label{eq18}
\begin{aligned}[b]
\mathcal{F}_{\{n_L\}}[E,x(\tau)]&=\int_{-\infty}^{+\infty}
\int_{-\infty}^{+\infty}e^{-iE(\tau_1-\tau_2)}\\
&\times\bra{\{n_L\}}\varphi[x(\tau_1)]\,\varphi[x(\tau_2)]
\ket{\{n_L\}}\,d\tau_1\,d\tau_2\ ,
\end{aligned}
\end{equation}
where $x(\tau)$ denotes the coordinates of the detector's worldline at proper
time~$\tau$.  For a detector comoving in the geometry of metric~\eqref{eq01},
at angular coordinates~$\Omega_N$, this too decomposes into a vacuum
contribution plus contributions from each excitation:
\begin{subequations}
\label{eq19}
\begin{equation}
\label{eq19a}
\begin{aligned}[b]
\mathcal{F}_{\{n_L\}}(E,\Omega_N)&=\mathcal{ F}_{\{0\}}(E,\Omega_N)\\
+a^{-N}&\sum_Ln_L\bigl[\abst{X_L(E)}^2+\abst{X_L(-E)}^2\bigr]
\mathcal{Y}_L^2(\Omega_N)\ ,
\end{aligned}
\end{equation}
with $X_L(E)$ the Fourier transform
\begin{align}
\nonumber
X_L(E)&=\int_{-\infty}^{+\infty}e^{-iEt}\chi_L(t)\,dt\\*
\nonumber
&=\frac{2^{(N-2)/2}a}{(2q)^{1/2}\Gamma(N/2)}\\*
\label{eq19b}
&\qquad\times[\kappa_L^{(+)}\Xi(+q,\lambda_1,E)
+\kappa_L^{(-)}\Xi(-q,\lambda_1,E)]\\
\intertext{and}
\nonumber
\Xi(q,\lambda_1,E)&=\left(\frac{\Gamma(1-iqa)}{\Gamma(1+iqa)}\right)^{1/2}
\Gamma\left(\frac{N}{4}+\frac{i(E+q)a}{2}\right)\\*
\nonumber
&\qquad\qquad\times\Gamma\left(\frac{N}{4}-\frac{i(E+q)a}{2}\right)\\*
\label{eq19c}
\times{}_3F_2\biggl(\frac{1}{2}-&\lambda_1,\,\frac{1}{2}+\lambda_1,\,
\frac{N}{4}+\frac{i(E+q)a}{2};\,1+iqa,\,\frac{N}{2};\,1\biggr)\ .
\end{align}
\end{subequations}
In Minkowski space the transform~$X$ is a delta function in energy.
The response function~$\mathcal{F}$ thus contains a factor enforcing energy
conservation, and a factor of the total time interval, by which it is divided
to give a response rate.  In de~Sitter space, however, the response probability
associated with an excitation is finite and has finite width in energy,
reflecting the influence of the dynamic spacetime geometry.

The choice of coefficients~$\kappa_L^{(\pm)}$ appearing in Eqs.~\eqref{eq06a},
\eqref{eq17}, and~\eqref{eq19b} is tantamount to the choice of vacuum state
on which the state space of the field is based.  Many authors have examined
the variety of vacuum states available for a scalar field in de~Sitter
space~\cite{burg84}.  For many reasons, the Euclidean~\cite{gibb77} or
Chernikov-Tagirov~\cite{cher68} vacuum emerges as the most appropriate
choice~\cite{redm89b}.  The coefficients corresponding to that choice are
\begin{subequations}
\label{eq20}
\begin{align}
\nonumber
\kappa_L^{(+)}&=\frac{2^{iqa}e^{-i\pi(\lambda_1-1/2)/2}}{
(1-e^{-2\pi qa})^{1/2}}\\*
\label{eq20a}
&\quad\times\left(\frac{\displaystyle{\Gamma\left(\frac{\frac{3}{2}
+\lambda_1+iqa}{2}\right)\,\Gamma\left(\frac{\frac{1}{2}+\lambda_1+iqa}{2}
\right)}}{\displaystyle{\Gamma\left(\frac{\frac{3}{2}+\lambda_1-iqa}{2}\right)
\,\Gamma\left(\frac{\frac{1}{2}+\lambda_1-iqa}{2}\right)}}\right)^{1/2}\\
\nonumber
\kappa_L^{(-)}&=\frac{2^{-iqa}e^{-i\pi(\lambda_1+3/2)/2}}{
(e^{2\pi qa}-1)^{1/2}}\\*
\label{eq20b}
&\quad\times\left(\frac{\displaystyle{\Gamma\left(\frac{\frac{3}{2}
+\lambda_1-iqa}{2}\right)\,\Gamma\left(\frac{\frac{1}{2}+\lambda_1-iqa}{2}
\right)}}{\displaystyle{\Gamma\left(\frac{\frac{3}{2}+\lambda_1+iqa}{2}\right)
\,\Gamma\left(\frac{\frac{1}{2}+\lambda_1+iqa}{2}\right)}}\right)^{1/2}\ .
\end{align}
\end{subequations}
This choice will be used here in all subsequent calculations.

\section{LOCALIZED PARTICLE STATES\label{sec3}}

\subsection{``Smeared'' one-particle states\label{sec3a}}

A simple way to associate an excited state of the field~$\varphi$ with
a localized ``wave packet'' is to ``smear'' the field operator~$\varphi$
with a suitable classical solution~$\Phi$ of the wave equation~\eqref{eq03},
and apply the resulting operator to the vacuum state~$\ket{\{0\}}$.  The
result is a superposition of one-particle states:
\begin{subequations}
\label{eq21}
\begin{equation}
\label{eq21a}
\ket{\Phi}_1=(\Phi,\varphi)\,\ket{\{0\}}\ ,
\end{equation}
where the smeared field operator is the Klein-Gordon inner product
\begin{equation}
\label{eq21b}
(\Phi,\varphi)=i\int\left(\Phi\,\dot{\varphi}-\dot{\Phi}\,\varphi\right)
\,a^N\cosh^N(t/a)\,d\Omega_N\ .
\end{equation}
\end{subequations}
If the classical wave packet~$\Phi$ is given by the expansion
\begin{equation}
\label{eq22}
\Phi(t,\Omega_N)=a^{-N/2}\sum_L[\xi_L\chi_L(t)+\xi_L^*\chi_L^*(t)]
\,\mathcal{ Y}_L(\Omega_N)\ ,
\end{equation}
with $c$-number coefficients~$\xi_L$, then operator expansion~\eqref{eq04}
and the orthonormality of the normal-mode basis functions imply
\begin{equation}
\label{eq23}
(\Phi,\varphi)=\sum_L(\xi_L^*b_L-\xi_Lb_L^\dagger)\ .
\end{equation}
The smeared one-particle state is then
\begin{equation}
\label{eq24}
\ket{\Phi}_1=-\sum_L\xi_L\,\ket{1_L}\ ,
\end{equation}
where the kets on the right denote excitation-number eigenstates with one
excitation in the $L$~mode and zero in all others.  The state
$\ket{\Phi}_1$ is normalized to unity if the normalization condition
\begin{equation}
\label{eq25}
\sum_L\abst{\xi_L}^2=1
\end{equation}
is imposed on~$\Phi$.  Because this state is
a superposition of single-excitation particle-number eigenstates, the
expectation value of the field~$\varphi$ in this state is zero, just as
for a single eigenstate.

The excitation energy of this state, i.e., the expectation value of the
field Hamiltonian above the vacuum value, follows from Eq.~\eqref{eq16a}.
It is
\begin{equation}
\label{eq26}
\begin{aligned}[b]
\mathcal{ E}_1(\{\xi_L\},t)&={}_1\bra{\Phi}H\ket{\Phi}_1
-\bra{\{0\}}H\ket{\{0\}}\hfil\\
&=\sum_L\abst{\xi_L}^2\,E_L(t)\ ,
\end{aligned}
\end{equation}
with the $E_L$ from Eq.~\eqref{eq16b}.

The detector response function for this state can be calculated as in
Eq.~\eqref{eq18}.  For a comoving detector, the result is
\begin{subequations}
\label{eq27}
\begin{equation}
\label{eq27a}
\begin{aligned}[b]
\mathcal{ F}_{\Phi}^{(1)}(E,\Omega_N)&=\mathcal{ F}_{\{0\}}(E,\Omega_N)\\
&\qquad+\abst{\tilde\Phi_+(E,\Omega_N)}^2+\abst{\tilde\Phi_+(-E,\Omega_N)}^2\ ,
\end{aligned}
\end{equation}
where
\begin{equation}
\label{eq27b}
\tilde\Phi_+(E,\Omega_N)\equiv a^{-N/2}\sum_L
\xi_L\,X_L(E)\,\mathcal{ Y}_L(\Omega_N)\ ,
\end{equation}
\end{subequations}
with $X_L$ from Eq.~\eqref{eq19b}, is the temporal Fourier transform of the
``positive frequency'' part of the wave packet~$\Phi$.

\subsection{States with nonzero $\langle\varphi\rangle$\label{sec3b}}

One feature which might be expected of a quantum-field-theoretic description
of a classical or ``first-quantized'' particle, but which does not appear
in the smeared one-particle state, is a nonzero expectation value for the
wave field~$\varphi$.  This can be obtained via an admixture of vacuum
and one-particle states.  For example, the state
\begin{equation}
\label{eq28}
\ket{\Phi}_\epsilon=\cos\epsilon\,\ket{\{0\}}-\sin\epsilon\,\ket{\Phi}_1
\end{equation}
has field expectation value
\begin{equation}
\label{eq29}
{}_\epsilon\bra{\Phi}\varphi\ket{\Phi}_\epsilon=\sin\epsilon\,
\cos\epsilon\,\Phi(t,\Omega_N)\ .
\end{equation}
The energy of this state above the vacuum is
\begin{equation}
\label{eq30}
\mathcal{E}_\epsilon(\{\xi_L\},t)=\sin^2\epsilon\,\mathcal{E}_1(\{\xi_L\},t)\ ,
\end{equation}
with $\mathcal{E}_1$ from Eq.~\eqref{eq26}.  The detector response function
for this state is
\begin{equation}
\label{eq31}
\begin{aligned}[b]
\mathcal{F}_{\Phi}^{(\epsilon)}(E,\Omega_N)&=\mathcal{F}_{\{0\}}(E,\Omega_N)\\
&+\sin^2\epsilon\, \bigl[\abst{\tilde{\Phi}_+(E,\Omega_N)}^2
+\abst{\tilde{\Phi}_+(-E,\Omega_N)}^2\bigr]\ ,
\end{aligned}
\end{equation}
with $\tilde{\Phi}_+$ again from Eq.~\eqref{eq27b}.

\subsection{Coherent states\label{sec3c}}

Perhaps the most appropriate description of a ``classical'' particle or
wave packet in quantum field theory is a coherent state or Glauber
state~\cite{schr26}, akin to those states embodying classical behavior in a
quantum harmonic oscillator.  These states can be constructed by applying
a ``displacement operator''~\cite{schu86,zhan90} to the field ground state:
\begin{subequations}
\label{eq32}
\begin{equation}
\label{eq32a}
\ket{\Phi}_c=\prod_LD_L(\xi_L)\,\ket{\{0\}}\ ,
\end{equation}
with
\begin{equation}
\label{eq32b}
D_L(\xi_L)\equiv\exp(\xi_Lb_L^\dagger-\xi_L^*b_L)\ .
\end{equation}
\end{subequations}
Clearly this state is a superposition of all particle-number eigenstates.
It can be written
\begin{equation}
\label{eq33}
\begin{aligned}[b]
\ket{\Phi}_c&=\prod_L\exp(-\half\abst{\xi_L}^2)\,
\exp(\xi_Lb_L^\dagger)\,\exp(-\xi_L^*b_L)\,\ket{0}_L\hfil\\
&=e^{-\half}\prod_L\left(\sum_{k=0}^\infty\frac{\xi_L^k}{\sqrt{k!}}\,
\ket{k}_L\right)\ .
\end{aligned}
\end{equation}
Here $\ket{k}_L$ denotes the $k$th excited state of the
$L$ field mode, and the normalization~\eqref{eq25} is again assumed.

An explicit wave functional for this state can be obtained by using
operator representation~\eqref{eq12}.  With the time dependence of each term
in expansion~\eqref{eq22} of~$\Phi$,
\begin{subequations}
\label{eq34}
\begin{equation}
\label{eq34a}
Y_L(t)\equiv a^{-1/2}[\xi_L\chi_L(t)+\xi_L^*\chi_L(t)]\ ,
\end{equation}
and a corresponding time derivative
\begin{equation}
\label{eq34b}
\mathcal{P}_L(t)\equiv a^{1/2}\cosh^N(t/a)\,[\xi_L\dot\chi_L(t)+
\xi_L^*\dot\chi_L^*(t)]\ ,
\end{equation}
the displacement operator takes the form
\begin{equation}
\label{eq34c}
\begin{aligned}[b]
D_L(\xi_L)&=\exp\left(i\mathcal{ P}_L\,y_L-Y_L\,\frac{\delta}{\delta y_L}
\right)\\
&=e^{-\frac{i}{2}\mathcal{ P}_LY_L}\,e^{i\mathcal{ P}_Ly_L}\,
e^{-Y_L\frac{\delta}{\delta y_L}}\ .
\end{aligned}
\end{equation}
\end{subequations}
Applying these operators to a wave functional of form~\eqref{eq13a} for
the vacuum state yields the wave functional
\begin{equation}
\label{eq35}
\begin{aligned}[b]
\Psi_{\Phi}^{(c)}[\{y_L\},t]&=\prod_L\left(\frac{a}{2\pi}\right)^{1/4}
[\chi_L^*(t)]^{-1/2}e^{-\frac{i}{2}\mathcal{ P}_LY_L}\,e^{i\mathcal{ P}_Ly_L}\\
&\quad\times\exp\{-\half[\Delta_L(t)-i\bar\Delta_L(t)]\,(y_L-Y_L)^2\}
\end{aligned}
\end{equation}
for the coherent state corresponding to the classical field~$\Phi$.

Expectation values of operators in this state match closely the corresponding
classical quantities.  The expectation value of the field is
\begin{equation}
\label{eq36}
{}_c\bra{\Phi}\varphi\ket{\Phi}_c=\Phi(t,\Omega_N)\ .
\end{equation}
The excitation energy of the state, above the vacuum, is
\begin{equation}
\label{eq37}
\begin{aligned}[b]
\mathcal{E}_c(\{\xi_L\},t)=\sum_L&\left(\frac{\sech^N(t/a)}{2a}\,
\mathcal{P}_L^2(t)\right.\\
&\qquad\quad+\left.\frac{a\cosh^N(t/a)}{2}\,\omega_L^2(t)\,Y_L^2(t)\right)\ .
\end{aligned}
\end{equation}
The operator identity
\begin{subequations}
\label{eq38}
\begin{equation}
\label{eq38a}
e^A\,B\,e^{-A}=\sum_{n=0}^\infty\frac{1}{n!}[A,B]^{(n)}\ ,
\end{equation}
with
\begin{align}
\label{eq38b}
[A,B]^{(0)}&\equiv B\\
\intertext{and}
\label{eq38c}
[A,B]^{(n)}&\equiv\bigl[A,[A,B]^{(n-1)}\bigr]\ ,
\end{align}
\end{subequations}
as may be proved by induction, implies, e.g.,
\begin{subequations}
\label{eq39}
\begin{align}
\label{eq39a}
b_L\,D_L(\xi_L)&=D_L(\xi_L)\,(b_L+\xi_L)\\
\intertext{and}
\label{eq39b}
[D_L^(\xi_L)]^\dagger\,b_L^\dagger&=(b_L^\dagger+\xi_L^*)\,[D_L(\xi_L)]^\dagger
\end{align}
\end{subequations}
These can be used to evaluate the comoving-detector response
function in the coherent state, yielding
\begin{subequations}
\label{eq40}
\begin{equation}
\label{eq40a}
\begin{aligned}[b]
\mathcal{F}_c(E,\Omega_N)&=\mathcal{F}_{\{0\}}(E,\Omega_N)
+\abst{\tilde{\Phi}(E,\Omega_N)}^2\\
&=\mathcal{F}_{\{0\}}(E,\Omega_N)\\
&\qquad+[\tilde{\Phi}_+(E,\Omega_N)+\tilde{\Phi}_+^*(-E,\Omega_N)]\\
&\qquad\qquad\times[\tilde{\Phi}_+(-E,\Omega_N)+\tilde{\Phi}_+^*(E,\Omega_N)]\\
&=\mathcal{F}_\Phi^{(1)}(E,\Omega_N)\\
&\qquad+2\Re{[\tilde{\Phi}_+(E,\Omega_N)\tilde{\Phi}_+(-E,\Omega_N)]}\ ,
\end{aligned}
\end{equation}
with
\begin{equation}
\label{eq40b}
\tilde{\Phi}(E,\Omega_N)\equiv\int_{-\infty}^\infty e^{-iEt}\,
\Phi(t,\Omega_N)\,dt
\end{equation}
\end{subequations}
the temporal Fourier transform of the \emph{full} wave packet~$\Phi$,
and $\mathcal{F}_\Phi^{(1)}$ and $\tilde{\Phi}_+(E,\Omega_N)$ from
Eqs.~\eqref{eq27}.  Result~\eqref{eq40a} shows that as long as
$\Re{[\tilde{\Phi}_+(E,\Omega_N)\tilde{\Phi}_+(-E,\Omega_N)]}$ is nonzero,
an Unruh-DeWitt detector can distinguish between the smeared one-particle
state and the coherent field state for the same classical field
configuration~$\Phi$.

\section{EXAMPLE:  A GAUSSIAN WAVE PACKET\label{sec4}}

To illustrate the persistence of spacetime-dynamical effects on a localized
particle, I use a simple---if somewhat unrealistic---classical wave packet:
a momentarily static, spherically symmetric Gaussian field at the instant
of minimal spatial radius ($t=0$), centered on the "north pole"
($\theta_1=0$) of the de~Sitter space.  This is specified via
\begin{subequations}
\label{eq41}
\begin{align}
\label{eq41a}
\Phi(0,\Omega_N)&=\frac{K}{(2\pi\sigma^2)^{1/2}}\,
\exp\left(-\frac{\theta_1^2}{2\sigma^2}\right)\\
\intertext{and}
\label{eq41b}
\dot{\Phi}(0,\Omega_N)&=0\ ,
\end{align}
\end{subequations}
where constant~$\sigma$ represents the width of the initial Gaussian and
$K$ is determined numerically by enforcing condition~\eqref{eq25}.  A
separate normalization of the form $[a\cosh(t/a)]^N\int\Phi^2\,d\Omega_N$
is not imposed, as the field equation~\eqref{eq03} does not preserve such a
condition.

For numerical calculations, it is convenient to write the time-dependence
functions~\eqref{eq06a} in the equivalent form
\begin{subequations}
\label{eq42}
\begin{equation}
\label{eq42a}
\chi_L(t)=\left(\frac{a}{2\Re{\gamma_L}}\right)^{1/2}\left(f_1(t)
-i\frac{\gamma_L}{a}f_2(t)\right)\ ,
\end{equation}
with
\begin{equation}
\label{eq42b}
\begin{aligned}[b]
f_1(t)&=[\cosh(t/a)]^{-N/2-iqa}\\
\times&F\left(\frac{\half+\lambda_1+iqa}{2},
\frac{\half-\lambda_1+iqa}{2};\half;\tanh^2(t/a)\right)
\end{aligned}
\end{equation}
and
\begin{equation}
\label{eq42c}
\begin{aligned}[b]
f_2(t)&=[\cosh(t/a)]^{-N/2-iqa}\\
\times&F\left(\frac{\thrhlf+\lambda_1+iqa}{2},
\frac{\thrhlf-\lambda_1+iqa}{2};\thrhlf;\tanh^2(t/a)\right)\ ,
\end{aligned}
\end{equation}
\end{subequations}
the $F$'s here denoting ordinary hypergeometric functions.  The parameters
\begin{subequations}
\label{eq43}
\begin{equation}
\label{eq43a}
\gamma_L=\Delta_L(0)-i\bar{\Delta}_L(0)=ia\frac{\dot{\chi}_L(0)}{\chi_L(0)}\ ,
\end{equation}
with~$\Delta_L$ and~$\bar{\Delta}_L$ from Eqs.~\eqref{eq13},
incorporate the choice of vacuum state for the field, just as do the
coefficients~$\kappa_L^{(\pm)}$ of Eq.~\eqref{eq06a}.  For the Euclidean
or Chernikov-Tagirov vacuum choice used here, their values are
\begin{equation}
\label{eq43b}
\gamma_L=2\absd{\frac{\Gamma\left(\dfrac{\thrhlf+\lambda_1+iqa}{2}\right)}
{\Gamma\left(\dfrac{\half+\lambda_1-iqa}{2}\right)}}^2\ .
\end{equation}
\end{subequations}
This specification is equivalent to the coefficient choice~\eqref{eq20}.

All the particle properties of interest can be calculated once the
expansion coefficients~$\xi_L$ for the wave packet~$\Phi$ are known.
With the $\gamma_L$ real, condition~\eqref{eq41b} implies $\xi_L=\xi_L^*$.
Expansion~\eqref{eq22} and form~\eqref{eq42} then yield the general form
\begin{subequations}
\label{eq44}
\begin{equation}
\label{eq44a}
\xi_L=a^{(N-1)/2}\left(\frac{\gamma_L}{2}\right)^{1/2}
\int\Phi(0,\Omega_N)\,\mathcal{Y}_L(\Omega_N)\,d\Omega_N\ .
\end{equation}
The calculations displayed here are for the case of three spatial
dimensions ($N=3$), and a value for parameter~\eqref{eq06b} of
$qa=\half$.  The spherical symmetry of the wave packet implies that
only $\ell_2=\ell_3=0$ harmonics contribute.  The relevant coefficients
are thus
\begin{equation}
\label{eq44b}
\begin{aligned}[b]
\xi_{\ell_100}&=4\pi a\left(\frac{\gamma_{\ell_1}}{2}\right)^{1/2}
\frac{K}{(2\pi\sigma^2)^{1/2}}\\
&\qquad\times\int_0^\pi\mathcal{Y}_{\ell_100}(\theta_1)
\,\exp\left(-\frac{\theta_1^2}{2\sigma^2}\right)\,\sin^2\theta_1\,d\theta_1\\
&=\frac{2aK}{\sigma}(\pi\gamma_{\ell_1})^{1/2}\\
&\times\int_0^\pi\frac{1}{(2\pi^2)^{1/2}}\frac{\sin(\lambda_1\theta_1)}
{\sin\theta_1}\,\exp\left(-\frac{\theta_1^2}{2\sigma^2}\right)\,
\sin^2\theta_1\,d\theta_1\\
&\cong\frac{aK\gamma_{\ell_1}^{1/2}}{2}\left(e^{-\ell_1^2\sigma^2/2}
-e^{-(\ell_1+2)^2\sigma^2/2}\right)\ .
\end{aligned}
\end{equation}
The approximation in the last expression is that of extending the integration
from~$\pi$ to infinity; for $\sigma$~values substantially less than unity
the error introduced is negligible.  It is even possible to calculate the
coefficients~$\xi_{\ell_100}$ for the $\sigma\to0$ limit of this wave
packet, i.e., a genuinely pointlike $\delta$-function packet at $t=0$.
This yields
\begin{equation}
\label{eq44c}
\xi_{\ell_100}^{(\delta)}=\frac{aK}{2\pi}\gamma_{\ell_1}^{1/2}(\ell_1+1)\ ,
\end{equation}
\end{subequations}
where again $K$ is determined numerically by imposing condition~\eqref{eq25}
on the coefficients.  In this case, since the expansion for~$\Phi$ is
formally divergent, care must be taken to discard numerical artifacts
associated with the cutoff in the sum over $\ell_1$~values.  Here the
convergent series for finite $\sigma$~values are taken to 50~terms; the
series for the $\sigma\to0$ limit are cut off at 100~terms.

The classical evolution of the wave packet, and the associated quantum energies
and detector responses described in the previous section, are obtained via
numerical evaluation of the necessary gamma and hypergeometric functions.
As both of these can involve sums of large terms with alternating signs,
roundoff errors are a significant concern.  The results shown here were
calculated in quadruple-precision---128-bit real and 256-bit
complex---arithmetic to suppress these errors.  The sums over~$\ell_1$
were taken over 50~terms for the (convergent) finite-$\sigma$ cases,
and over 100~terms for the $\sigma\to0$ limit case.

\begin{figure*}
\includegraphics{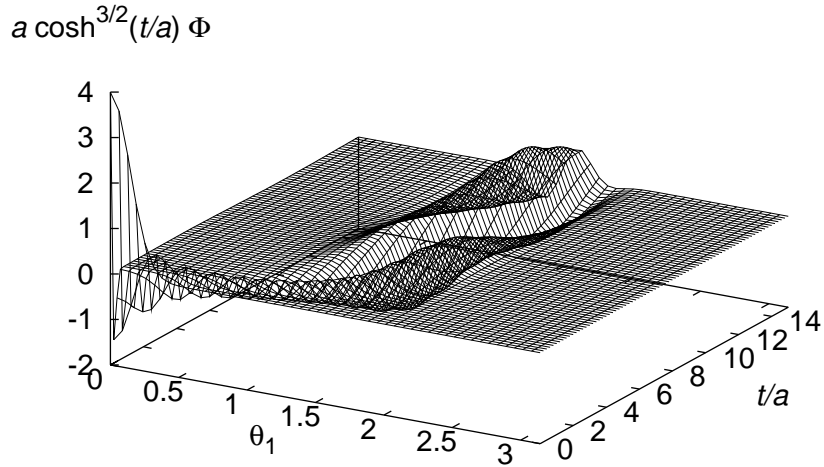}
\caption{\label{f1}Evolution of a spherically symmetric, initially Gaussian
scalar wave packet in de~Sitter space, centered on the ``north pole''
($\theta_1=0$).  The (angular) width of the initial Gaussian here is
$\sigma\doteq0.1414\ .$  The field~$\Phi$ is rescaled by $\cosh^{3/2}(t/a)$
to compensate for the physical expansion of the space.}
\end{figure*}

The classical evolution of the Gaussian wave packet is shown in
Fig.~\ref{f1}.
The field~$\Phi$ is shown rescaled by a factor~$\cosh^{3/2}(t/a)$, to
compensate for the diminution of~$\Phi$ associated with the increasing
physical volume of the space.  Over the time interval shown, the radius of
the space increases by a factor of over~$1.6\times10^6$, the volume by
over~$4.3\times10^{18}$.  The initially localized packet spreads quickly to the
two-dimensional ``equator'' of the space, i.e., the sphere $\theta_1=\pi/2$.
This general behavior is quite insensitive to the initial width~$\sigma$ of
the packet.

\begin{figure*}
\includegraphics{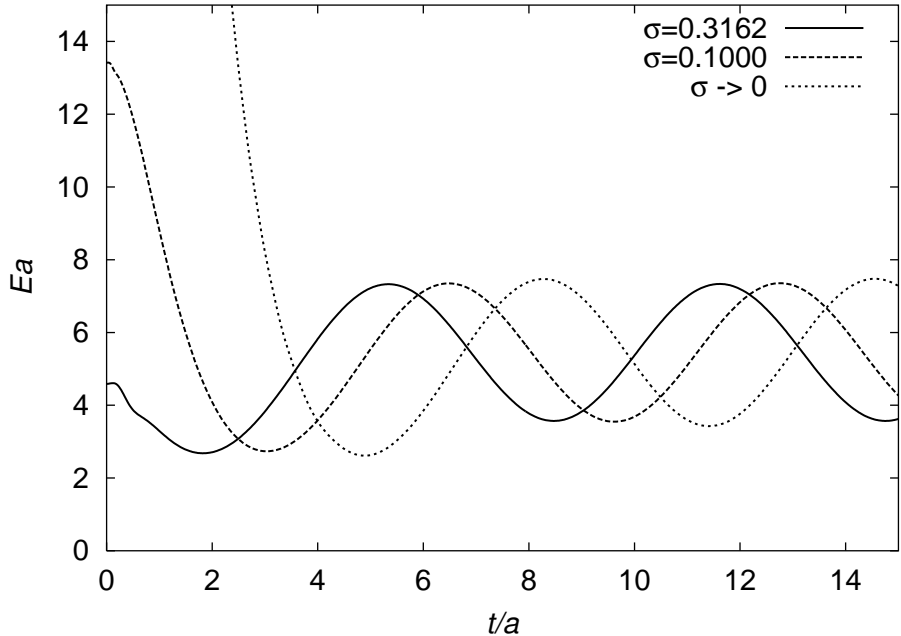}
\caption{\label{f2}Excitation energies $\mathcal{E}_c$ of coherent (Glauber)
field states corresponding to initially-Gaussian classical wave packets.
The energy for the initially pointlike ($\sigma\to0$) wave packet decreases
monotonically from a value of $80.4133\,a^{-1}$ at $t=0$ to the region
shown here.}
\end{figure*}

The excitation energies~\eqref{eq37} of coherent quantum field states
corresponding to such classical wave packets, above the Euclidean vacuum,
are shown in Fig.~\ref{f2}.
For the initially static wave packets used here, the energies are
time-symmetric.  As it happens, the excitation energies~\eqref{eq26} of
smeared one-particle states for the same wave packets are indistinguishable
from these on the scale of this graph.  The most striking features of these
energies are that the oscillations apparent in the individual normal-mode
energies persist, and the amplitudes of these oscillations---though not their
phases---are insensitive to the initial width of the classical wave packet.
Moreover, for the parameters used here, the mass of the field is
$\mu=1.581~a^{-1}$; the oscillating energies are always substantially above
that value.  This contrasts with the energy of a \emph{classical} particle,
which would rapidly redshift to the value~$\mu$.  The periods of the
oscillations all approach $6.28\,a$, the same as that of the
individual-mode energy oscillations~\eqref{eq17a} for the value of~$q$
taken here.

\begin{figure*}
\includegraphics{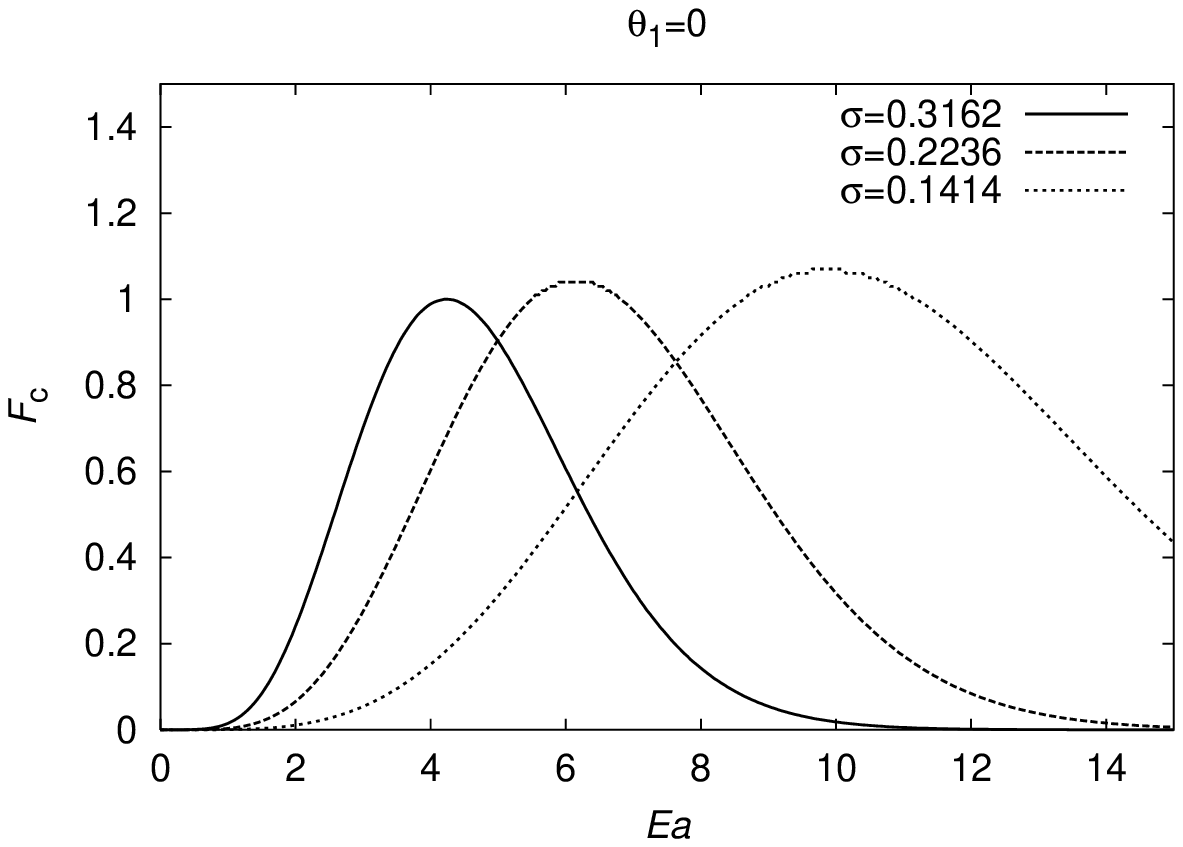}
\caption{\label{f3}Detector response functions $\mathcal{F}_c$ for coherent
field states corresponding to initially-Gaussian wave packets, above the
Euclidean-vacuum contribution (``dark current'').  The detector is comoving
at the ``north pole'' $\theta_1=0$.  Responses for smeared one-particle states
corresponding to the same wave packets are indistinguishable from these on
this scale.}
\end{figure*}

\begin{figure*}
\includegraphics{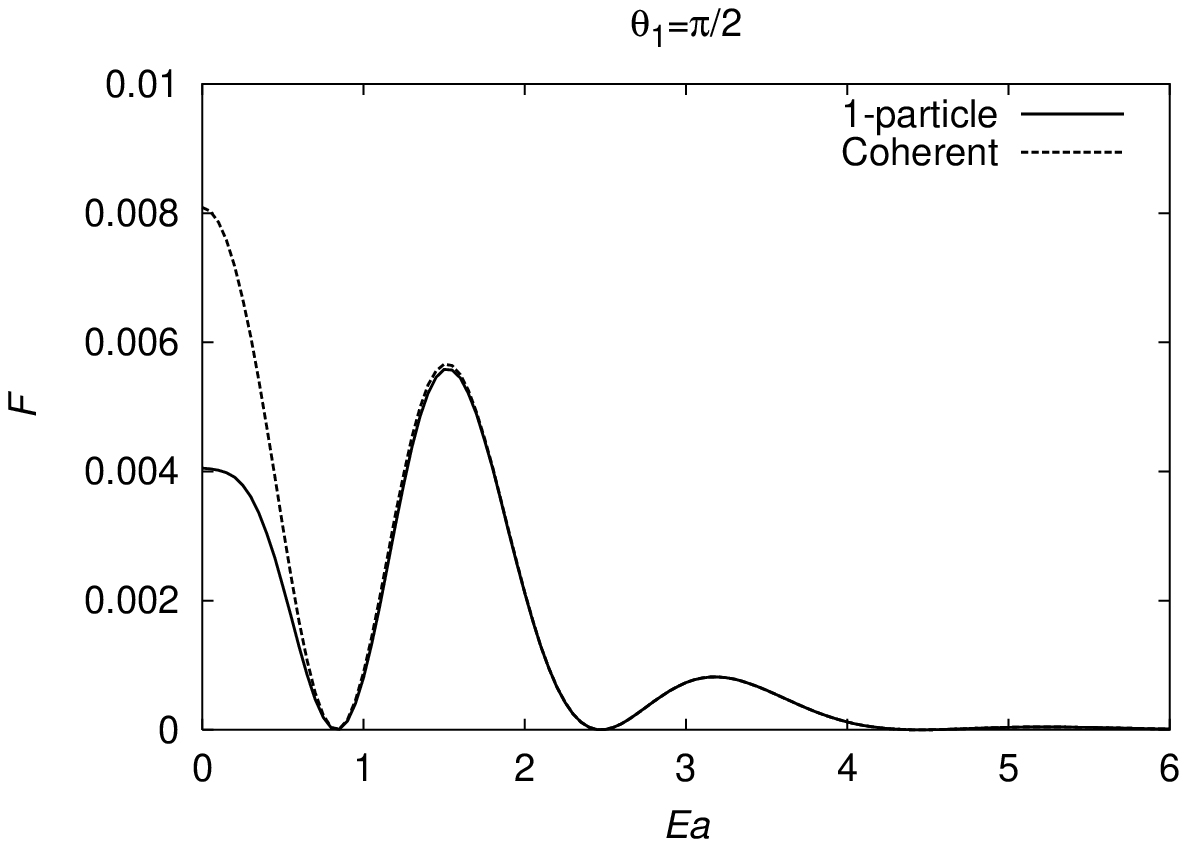}
\caption{\label{f4}Detector response functions $\mathcal{F}_1$ and
$\mathcal{F}_c$ for smeared one-particle and coherent states corresponding
to an initially-Gaussian wave packet with $\sigma=0.2236$. Here the
detector is comoving at the ``equator'' ($\theta_1=\pi/2$).}
\end{figure*}

\begin{figure*}
\includegraphics{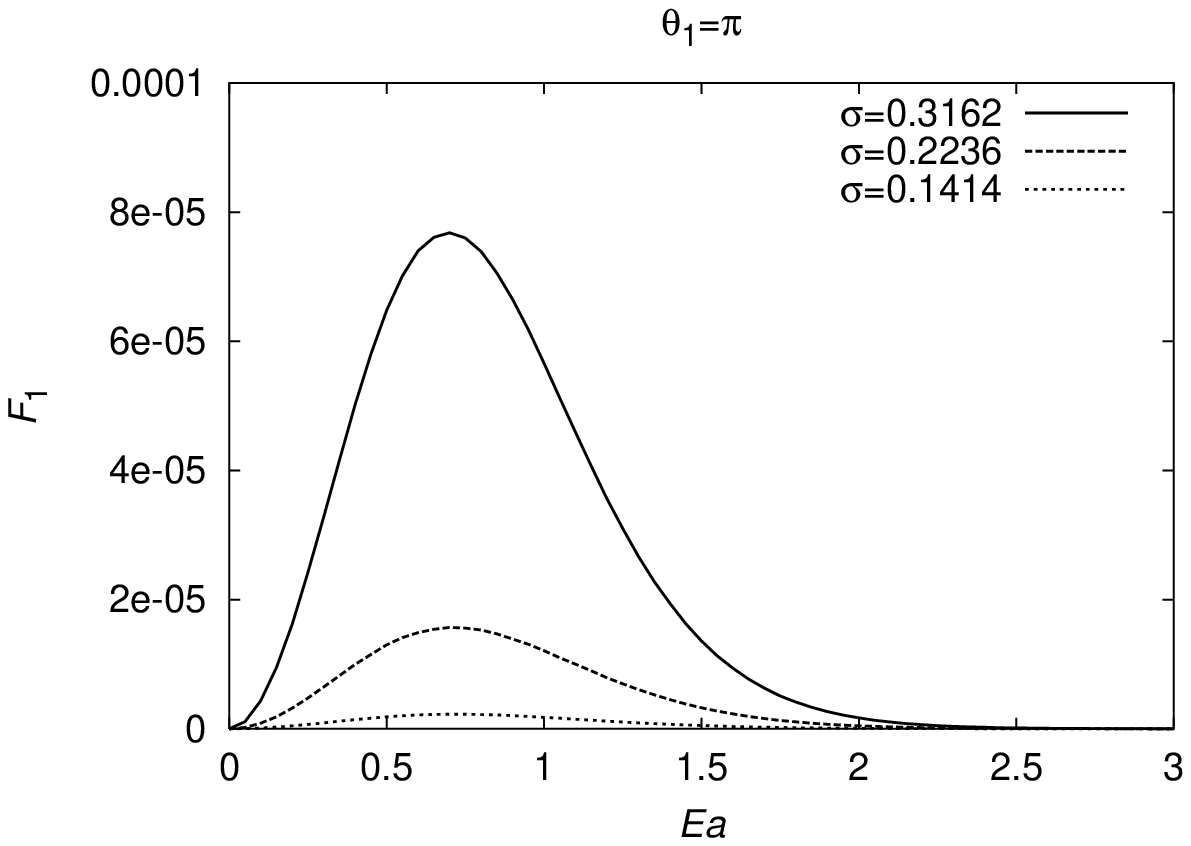}
\caption{\label{f5}Detector responses $\mathcal{F}_1$ for smeared one-particle
states corresponding to initially-Gaussian wave packets.  Here the detector
is comoving at the ``south pole'' ($\theta_1=\pi$).  Responses for coherent
states for the same wave packets, for this detector position, are negligible
on the scale of this graph (and below the resolution of the calculation).}
\end{figure*}

The detector response functions, above the vacuum contribution, for
smeared one-particle and coherent field states corresponding to these wave
packets are shown in Figs.~\ref{f3}--\ref{f5}.  Because the wave packets are
localized, the response depends strongly on the location of the detector.
Responses for coherent states, with the detector comoving---i.e., sitting---at
the ``north pole'' of the space ($\theta_1=0$, the center of the wave packet)
are shown in Fig.~\ref{f3}.  Response functions for smeared one-particle states
corresponding to the same wave packets are indistinguishable from these at
this position.
However, for a detector comoving anywhere on the ``equator''
($\theta_1=\pi/2$), the responses for the two field states are distinguishable
at low energies, although the response functions (probabilities) are some
two orders of magnitude smaller.  This is illustrated in Fig.~\ref{f4} for
a wave packet with $\sigma=0.2236$; the results for other widths are similar
in size, shape, and location.
With the detector at the ``south pole'' ($\theta_1=\pi$), the responses for
smeared one-particle states are two orders of magnitude smaller still, as
shown in Fig.~\ref{f5}.  The responses for coherent states for the same wave
packets are at least three orders of magnitude smaller even than these at this
position; in fact they are below the precision of these calculations.
In all cases the response functions retain the finite peak height and width
which characterize the individual normal-mode excitations.  Of course here
both the irreducible, individual-mode widths and the superposition of modes
required to form the wave packets contribute to the observed peak widths.
The peak energies increase monotonically with increasing localization of
the wave packet (decreasing~$\sigma$), as might be expected, for the
north-pole detector (Fig.~\ref{f3}).  Notably, however, this progression is
not apparent at the other detector positions.  There the peak \emph{heights}
decrease monotonically with decreasing~$\sigma$, i.e., the more strongly
localized particles are simply less detectable at these locations. 

\section{CONCLUSIONS\label{sec5}}

Localized first-quantized or classical particles are readily described in
a quantized linear field theory on a curved-spacetime background.  Smeared
one-particle field states and coherent field states provide straightforward
means for this.  Dynamical features of the particles, such as energy
expectation values and detector response functions in these states, are
easily evaluated.  For particles in hyperbolically (exponentially) expanding
de~Sitter space, the distinctive behavior of these features for normal-mode
excitations of the field persists for at least some localized-particle
states:  Energy expectation values oscillate, and detector response functions
are finite, with finite width in energy.  While states describing more
realistic particles, e.g., those localized at arbitrary positions at
arbitrary times, would require more complicated normal-mode expansions
than those used here, it is to be expected that these behaviors would
still characterize the particles.

The smeared one-particle state construction is more commonly associated with
a single particle in quantum field theory, although coherent field states
might be more apt descriptions of a particle with a well-defined
(first-quantized) wave function, or of a classical field wave packet.
For the simple wave packets examined here, the energy expectation values
of the states do not distinguish between the two descriptions, but the
detector response functions---especially, for detectors located away from
the initial localization of the wave packets---could do so. 

The interpretation of these features in terms of physical measurements
is, however, somewhat subtle.  For example, if the scale factor~$a$ in
de~Sitter metric~\eqref{eq01} is taken to be $c/H_0$, with $c$ the speed
of light and $H_0=75~\hbox{km/s/Mpc}$, then its value in ordinary units
is $a\doteq1.2\times10^{26}$~m.  The field mass~$\mu$ corresponding
to the examples of Sec.~\ref{sec4} above (with $qa=\half$) is
$\mu\doteq1.58\hbar c/a\doteq2.5\times10^{-33}$~eV.  This is of
the same order of magnitude as the ultralow mass proposed by Parker
and Raval~\cite{park99a,park99b}.  But in that case the period of the
energy oscillations of Fig.~\ref{f2} is $2\pi a/c\doteq13$~Gyr---nearly
the presently accepted age of the Universe.  Larger $q$ and $\mu$ values
yield faster oscillations but much smaller amplitudes.  And for the energy
oscillations to be detectable, $\mathcal{E}$ must be measurable to a
precision substantially smaller than the oscillation amplitude in a time
smaller than, say, one radian of the oscillation.  With the values used here
the late-time amplitudes are all about~$2\,a^{-1}$, while the time for one
radian of oscillation is $a$ (in geometrized units once again), so energy-time
uncertainty allows detection, but with little leeway.  Of course these energies
are not energy eigenvalues but expectation values, so measurements of large
numbers of identical particles may allow greater precision.  Moreover, as
the energy-time uncertainly relation arises not from fundamental commutation
relations but from analysis of the measurement interaction, a more detailed
examination of the energy-measurement process might be needed to clarify this
issue.  In this respect the detector responses in Figs.~\ref{f3}--\ref{f5}
could be considered prototypical particle-energy measurements.  These are
integrals over all time, so they reveal no time dependence.  (Switching
the detectors on and off introduces extraneous excitations, difficult to
disentangle from interactions with the field~$\varphi$.)  And the energy
dependence of the responses is sensitive to detector location.  In these
calculations the detector is taken to be a pointlike quantum system
unaffected by the spacetime geometry---hence the ordinary Fourier transforms
in Eq.~\eqref{eq18}.  A future analysis of the interaction of two
genuinely de~Sitter-space quantum systems may shed more light on the
prospects for observing such quantum/spacetime-dynamical phenomena.  

\begin{acknowledgments}
Portions of this work were supported by the U.~S.~National Science
Foundation under Grants No.~PHY89--22140 at Washington University in St.~Louis
and No.~PHY91--05935 at the University of Wisconsin--Milwaukee.  I thank
Professor Robert Pasken of the Department of Earth and Atmospheric
Sciences, Saint Louis University, for invaluable assistance with the
numerical calculations of Sec.~\ref{sec4}.
\end{acknowledgments}


\end{document}